\documentclass[a4paper, 10pt, twocolumn]{article}
\pdfoutput=1 

\usepackage{graphicx}
\usepackage[noblocks]{authblk}
\usepackage{url}
\makeatletter
\g@addto@macro{\UrlBreaks}{\UrlOrds}
\makeatother
\usepackage{csquotes}

\newcommand{\Iote}{\enquote{IoT-enabled}}
\newcommand{\Iotd}{\enquote{IoT-disabled}}
\providecommand{\keywords}[1]{\noindent \textbf{Keywords:} #1}

\begin{document}

\title{\bf A city-scale IoT-enabled ridesharing platform}

\author[a,*]{\bf Claudio Gambella}
\author[a]{\bf Julien Monteil}
\author[a]{\bf Anton Dekusar}
\author[a]{\bf Sergio Cabrero Barros}
\author[a]{\bf Andrea Simonetto}
\author[a]{\bf Yassine Lassoued}
\affil[a]{\bf IBM Research, Ireland Lab, Mulhuddart, Dublin 15}
\affil[*]{Corresponding author email: \texttt{claudio.gambella1@ie.ibm.com}}

\date{}

\maketitle 

\begin{abstract}
\bf \quad The advent of on-demand mobility systems is expected to have a tremendous potential on the wellness of transportation users in cities. Yet such positive effects are reached when the systems under consideration enable seamless integration between data sources that involve a high number of transportation actors. In this paper we report on the effort of designing and deploying an integrated system, including algorithms and platforms, that can operate in cities, in an Internet of Things (IoT)-aware fashion. The system was evaluated by enabling/disabling the IoT components of the system, highlighting the necessity of real-time data integration for efficient mobility services.\\
\end{abstract}


\noindent\textbf{Acknowledgement:} This project has received funding from the European Union's Horizon 2020 research and innovation programme under grant agreement No 731993 (Autopilot).


\section{Introduction}
\label{sec-introduction}

Ridesharing consists in optimising the pickup and drop-off locations and routes of potentially-shared rides, using a fleet of vehicles. Ridesharing is typically solved in a centralised fashion across the whole vehicle fleet and across customers, as to minimise the total travel time of the operating vehicles, while taking into account the customer preferences (pickup and drop-off times, whether sharing is desired or not, etc.) \cite{simonetto2019real, pandey2019needs, alonso2017demand}. Ridesharing emerged in the past few decades amongst the shared-mobility services as a means to limit traffic congestion and achieve environmental benefits. In such a context, external information about the traffic situation and events affecting travel times, collected by IoT devices, may be taken into account in the scheduling of ridesharing. IoT is used to optimise the allocation, routes, pickup and drop-off locations based on real-time information received from the IoT ecosystem (e.g., traffic jams, incidents, road closures, traffic lights, etc.). While the effects of traffic conditions have been studied in small-sized ridesharing systems \cite{WangTL, XU2015161}, the inherent complexity of managing city-scale systems hindered their analysis on traffic-aware scenarios \cite{8317830}. In this paper, we propose a city-scale IoT implemented architecture and system to augment the optimisation-based ridesharing service proposed in \cite{simonetto2019real}. Such IoT-enabled ridesharing service has been developed as part of the Brainport pilot site of the European Horizon 2020 AUTOPILOT project.

\section{An IoT-Based Approach for Ridesharing}

\subsection{Architecture}

\begin{figure*}[hbt]
	\includegraphics[width=0.99\textwidth]{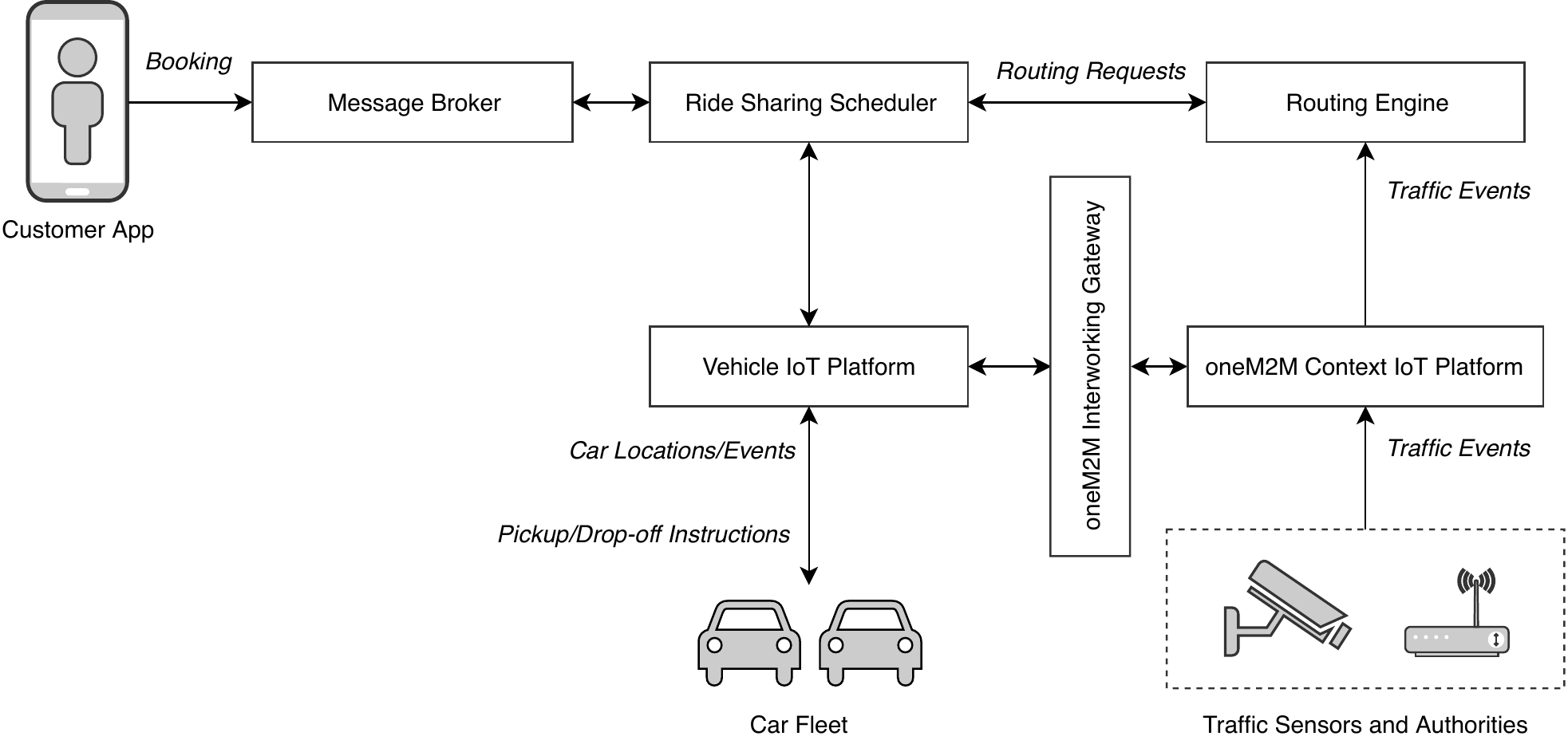}
	\caption{IoT-Based Ridesharing Service Architecture}
	\label{fig:arch}
\end{figure*}

Figure \ref{fig:arch} shows the architecture of the IoT-based ridesharing service that we implemented. As can be seen in this architectural diagram, a customer interacts with the service through a mobile app. Customer requests are communicated to the ride sharing service through a message broker and are processed in batches collected over regular time intervals. Shared cars communicate with the ridesharing service through a dedicated vehicle IoT platform. They act as {\em "things"}. Cars submit their events (locations, customer pickup and drop-off confirmations, detected events) to the vehicle IoT platform. They receive customer assignment and route instructions from the ridesharing scheduler through the same IoT platform. The ridesharing service queries a routing engine for the travel times between pairs of nodes of the area of interest. The routing engine receives notifications about changes in traffic speeds and situations through the oneM2M IoT platform, and updates its maps accordingly. In this way, traffic information is taken into account in the determination of travel times, and the assignment provided by the ridesharing scheduler takes this into account.

A standard oneM2M-based context IoT platform is used to access traffic events communicated by external devices, road sensors, and vehicles. The vehicle IoT platform exchanges data with the context IoT platform through a oneM2M interworking gateway. Data are exchanged as follows.
\begin{itemize}
\item In one direction, relevant traffic events affecting the operations of the shared car fleet are communicated from the oneM2M IoT platform to the vehicle IoT platform through the oneM2M interworking gateway. These may then be received by the shared cars to adjust their routes.
\item In the other direction, any events detected by the shared cars (e.g., accidents, hazards, etc.) are communicated from the vehicle IoT platform to the oneM2M IoT platform through the oneM2M interworking gateway.
\end{itemize}

\subsection{Ridesharing Optimisation}\label{sec-opt}

The aim of the ridesharing service is to find an assignment of available vehicles to customers so as to optimise a performance indicator, such as vehicle travel times. Candidate vehicles are those with residual seat capacity, located at an acceptable distance from the customer. Each customer specifies his trip request as an origin-destination coordinate pair and, optionally, time preferences for pickup and drop-off (e.g., maximum waiting time).
The ridesharing scheduler receives from a message broker a batch of customer requests at time interval $[t_{k-1}, t_k[$ and works as follows:
\begin{itemize}
	\item It obtains the customer requests submitted during the time interval $[t_{k-1}, t_k[$.
	\item It asks insertions costs to the vehicles indicated by the context mapping. This involves to solve a single-vehicle Dial-a-Ride problem (see, e.g., \cite{cordeau2003dial}), via a greedy insertion heuristic.
	\item It queries an optimisation module for an optimal assignment of vehicles to requests, which is modelled as a linear assignment problem (see, e.g., \cite{bertsekas1990auction}).
	\item If some customers cannot be served, it calls an internal rebalancing module, which runs the logic again (steps 2. to 3.) with looser customer time constraints and for idle vehicles only.
	\item It sends the assignments and their corresponding routes to the customers and vehicles. 
\end{itemize}

\section{Evaluation}

We validated our framework in two ways. Firstly, we ran large-scale (city-wide) simulations with virtual vehicles and customers. This allowed us to evaluate the benefits of the IoT-based approach performance in terms of quality of service. Secondly, we demonstrated the ridesharing service using 2 real automated vehicles. This allowed us to verify the feasibility of the approach. 

\subsection{Simulations}

\begin{figure}
	\includegraphics[width=0.5\textwidth]{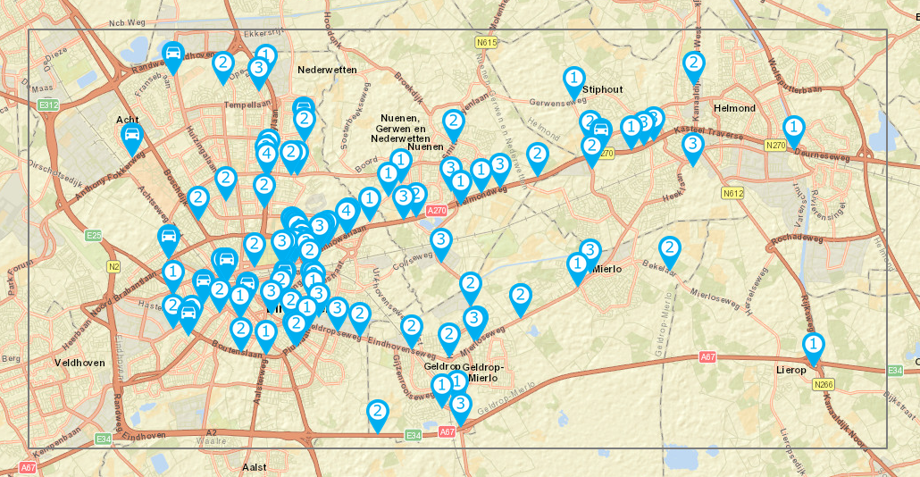}
	\caption{Screenshot of the simulation user interface showing the study area in Brainport, the Netherlands. Markers indicate the position of vehicles and the number of passengers they are carrying.}
	\label{fig:scenario}
\end{figure}

In this validation approach, we ran simulations in six regions in the Brainport area, Netherlands, with virtual fleets of cars and virtual customers. Here, we report on the Brainport simulations. As illustrated in the screenshot of Figure~\ref{fig:scenario}, the simulation area covers Eindhoven and Helmond and the highway that links both towns.
From these simulations, we draw conclusions relative to the scalability of our proposal and the advantages of integrating it with traffic data generated by IoT devices.

In the Brainport simulations, we generate a fleet of $100$ virtual vehicles serving between $360$ and $1440$ user requests during $1$ hour of service. Vehicles are assigned to requests every $10$ seconds: \cite{simonetto2019real} showed that such frequency for batch assignment provides an optimum trade-off between level of service and computation time. We have several speed cameras connected to the oneM2M IoT platform which provide traffic events every 5 min. These cameras are placed on a stretch of the highway between Eindhoven and Helmond. We recorded traffic events from the actual speed cameras for an hour and then we played them back during simulations. To make simulations more realistic we also simulated more speed cameras on the circular road around Eindhoven. Having gathered this IoT data we aim to evaluate its effect on the ridesharing system.

Each vehicle has the capacity to carry up to four passengers at the same time, and its location is initialised randomly in one of the six areas of interest of Eindhoven, Larger Eindhoven, Helmond, Neunen,  Geldrop, Whole area (grey box in Figure \ref{fig:scenario}), as reported in Table \ref{tab:prob-sim}. Users are simulated by generating ride requests, each consisting of the origin and destination coordinates for a single passenger. Passengers expect their requests to be attended within $7$ minutes. In case a request is not met, a rebalancing module attempts to serve it by assigning an idle vehicle as mentioned in the logic description in Section \ref{sec-opt}. 
 Similar to the initial vehicle locations, the origin of each request is calculated using six predefined areas with the probability indicated in Table \ref{tab:prob-sim}. Once the origin is selected, the destination is selected following the conditional probability documented in Table \ref{tab:prob-sim}. For example, if the origin is a random point in Larger Eindhoven, there is a $0.05$ probability of selecting a random point in Helmond as destination. Ride requests are grouped into batches of 10 seconds, and the batches are processed by the ridesharing service. Every batch has between $1$ and $4$ requests. 
For reproducibility purposes, ride requests are generated only once, offline, and then injected into the simulations at predefined time instants.

\begin{table*}[h!]
	\centering
	\begin{tabular}{c|cc|c|c|cccccc}
		& \multicolumn{2}{c|}{Bounding Box} & \multicolumn{1}{c|}{Vehicle} &	\multicolumn{1}{c|}{Origin}  & \multicolumn{6}{c}{Destination} \\
		& & &   &  & E & L & H & N & G & W \\
		\hline
E     &  51.4584, 5.5157  &  51.4154, 5.4453  & 0.30  & 0.30  & 0.30  & 0.30  & 0.05  & 0.15  & 0.15  & 0.05 \\
L     &  51.4927, 5.5323  &  51.4106, 5.4310  & 0.20  & 0.20  & 0.03  & 0.30  & 0.05  & 0.15  & 0.15  & 0.05 \\
H     &  51.5016, 5.7155  &  51.4511, 5.6008  & 0.20  & 0.20  & 0.10  & 0.10  & 0.60  & 0.05  & 0.05  & 0.10 \\
N     &  51.4814, 5.5756  &  51.4566, 5.5302  & 0.10  & 0.10  & 0.30  & 0.30  & 0.05  & 0.20  & 0.05  & 0.10 \\
G     &  51.4366, 5.5831  &  51.4068, 5.5342  & 0.10  & 0.10  & 0.30  & 0.30  & 0.05  & 0.05  & 0.20  & 0.10 \\
W     &  51.5021, 5.7249  &  51.4018, 5.3950  & 0.10  & 0.10  & 0.30  & 0.20  & 0.20  & 0.10  & 0.10  & 0.10 \\
	\end{tabular}
	\caption{Simulation parameters for the six areas of interest: Eindhoven (E), Larger Eindhoven (L), Helmond (H), Neunen (N),  Geldrop (G), Whole area (W, grey box in Figure \ref{fig:scenario}). Respectively, the columns indicate: the North East and South West coordinates of the bounding box of each area, the probability for the vehicles to have initial location in an area, the probability for the trips to have given origin and destination area.}
		\label{tab:prob-sim}
\end{table*}

The experiment setup consists of a custom traffic simulator that updates vehicle positions on a map considering the road network and traffic speeds per road link. The simulator is connected to the ride sharing service through the IoT platforms. The simulator is connected to the vehicle IoT platform to update vehicle routes and positions, and makes use of pre-recorded IoT event data sitting on the oneM2M IoT platform. So, from the service point of view, there is no difference between the simulator and a the real world. In addition, the simulator can be configured to simulate real-time traffic on selected roads, by updating the OSRM routing contract. We inserted IoT traffic data using this mechanism so as to simulate traffic jams in different roads. We are then able to use the routing engine in both the \Iote\ and \Iotd\ modes. In both modes, vehicles move at the appropriate speeds as measured by the IoT devices, and update their locations accordingly. The ridesharing scheduler is only aware of the IoT measurements in the \Iote\ condition. Thus, in the \Iotd\ condition, assignments are optimised assuming free flow traffic, i.e. traffic in all roads is fluid. The expectation is that this will lead the ridesharing scheduler to suboptimal recommendations that are detrimental to the service quality.

\begin{figure}[h!]
	\includegraphics[width=0.5\textwidth]{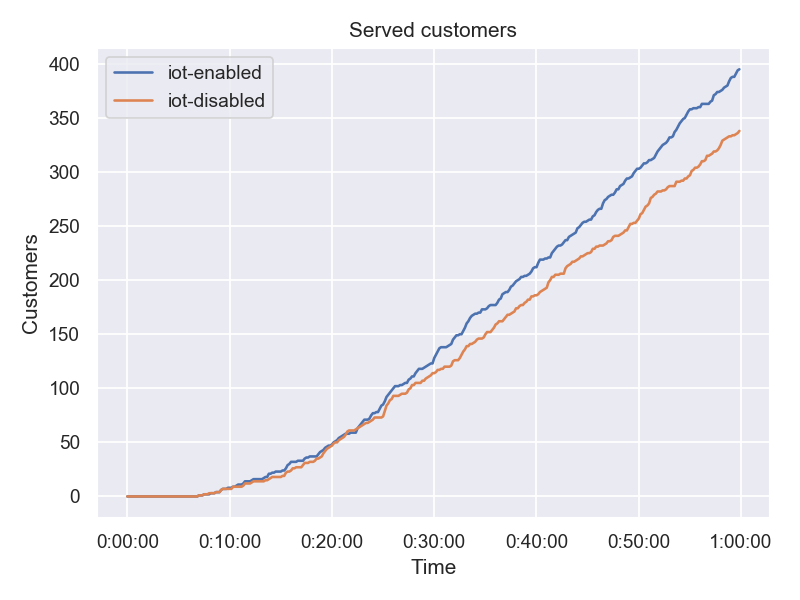}
	\caption{Number of users served over the one-hour simulation, for the \Iote\ and \Iotd\ ridesharing service.}
	\label{fig:customers-served}
\end{figure}

\begin{figure}[h!]
	\includegraphics[width=0.5\textwidth]{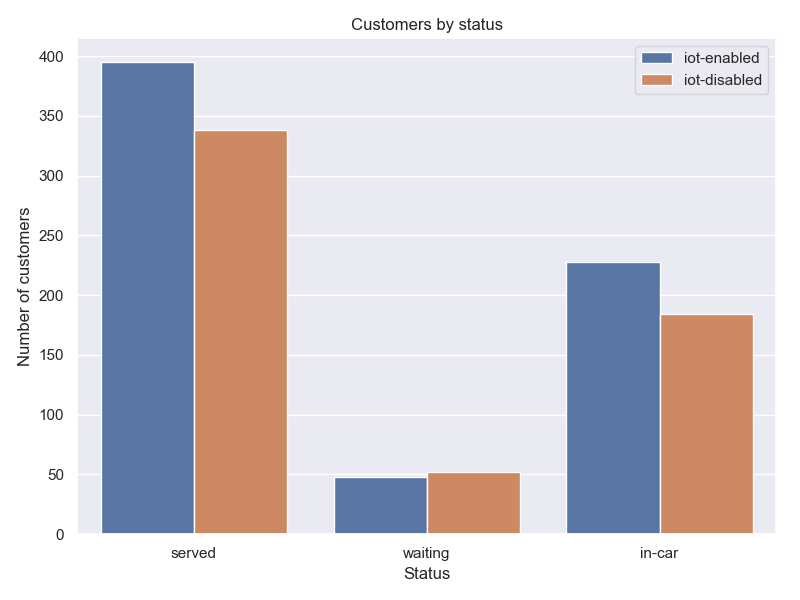}
	\caption{Customer status at the end of the one-hour simulation  for the \Iote\ and \Iotd\ ridesharing service. The histogram reports the number of served customers, the number of waiting customers, and the number of customers traveling on the cars.}
	\label{fig:customers-by-status}
\end{figure}

\begin{figure}[h!]
	\includegraphics[width=0.5\textwidth]{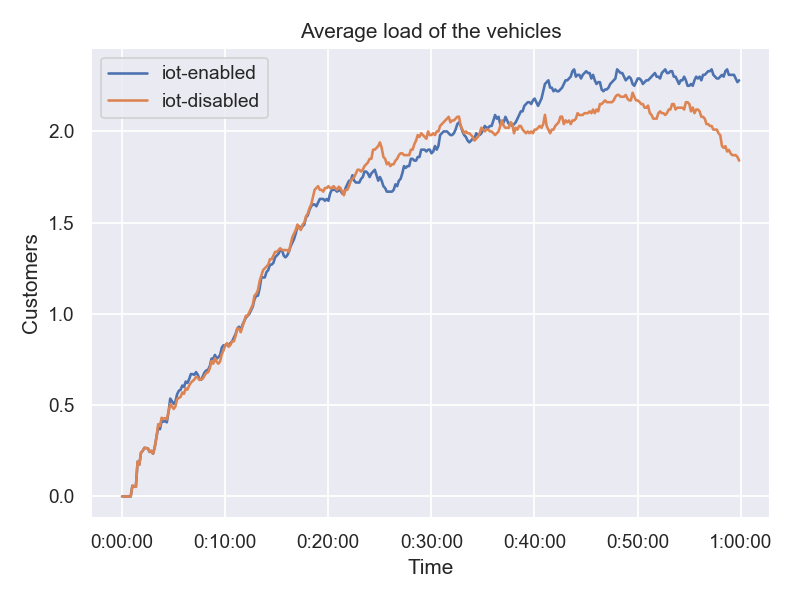}
	\caption{Average vehicle load during a one-hour simulation for the \Iote\ and \Iotd\ ridesharing service.}
	\label{fig:average-load}
\end{figure}

\begin{figure}[h!]
	\includegraphics[width=0.5\textwidth]{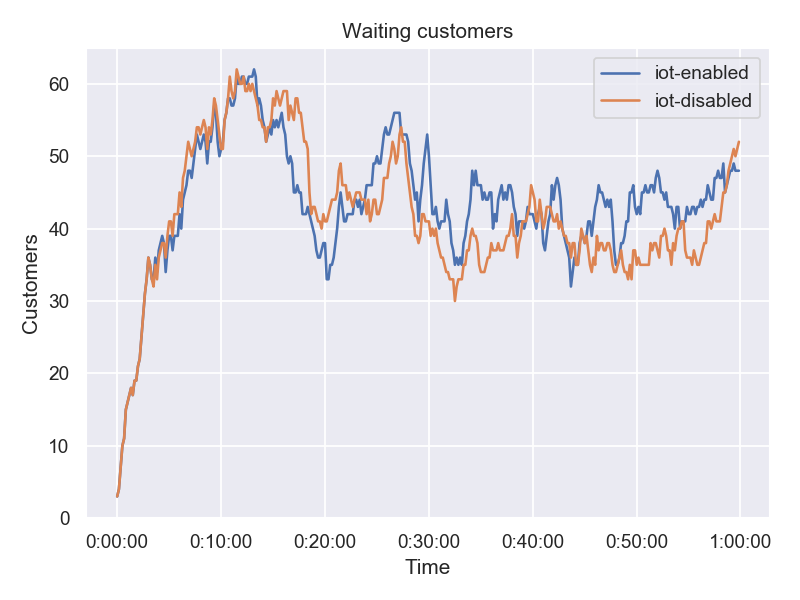}
	\caption{Number of customers waiting for a car during a one-hour simulation for the \Iote\ and \Iotd\ ridesharing service.}
	\label{fig:waiting}
\end{figure}

\begin{figure}[h!]
	\includegraphics[width=0.5\textwidth]{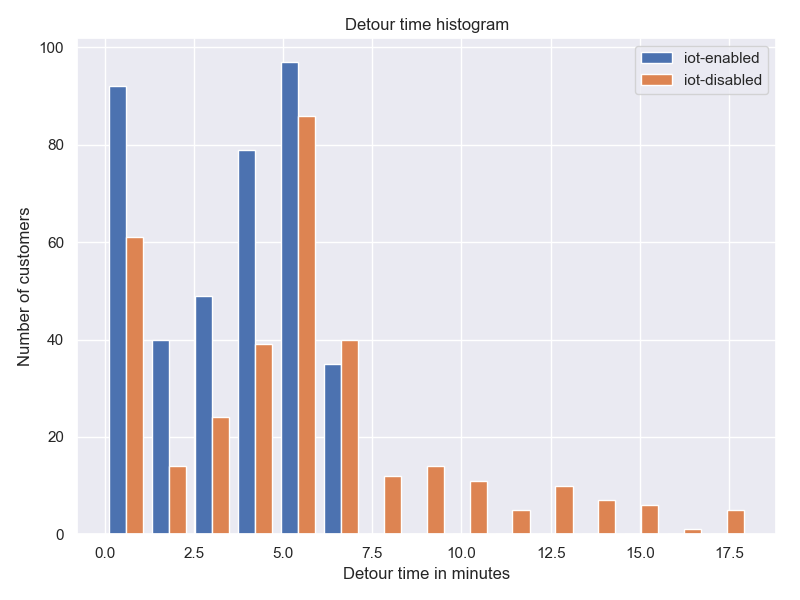}
	\caption{Cumulative detour time at the end of the one-hour simulation, for the \Iote\ and \Iotd\ ridesharing service.}
	\label{fig:detour-time-histogram}
\end{figure}

In the following, we show some results obtained for the area of interest and we compare the performance of the ridesharing service in both \Iote\ and \Iotd\ scenarios. Figure \ref{fig:customers-served} shows the evolution of the number of customers served over time, which is a common key performance indicator for ridesharing services (see, e. g., \cite{AGATZ2012295}). The growth of the served customers is larger for the \Iote\ service. Making ridesharing aware of IoT data has a positive impact not only on the number of users served, but also ion vehicle utilisation, and on the number of customers in waiting condition, as shown in Figure \ref{fig:customers-by-status}. 
By analysing the the average number of customers per vehicle in Figure \ref{fig:average-load} and the number of customers waiting for a ride in Figure \ref{fig:waiting}, it is possible to observe that the advantage of the \Iote\ service over the \Iotd\ one is noticeable towards the end of the simulation. This is because neglecting the variations of the travel times due to traffic appears to have a cumulative effect during the simulation. It is not possible however to establish a strict dominance of the \Iote\ service at each instant of the simulation for each of these indicators, as shown by the waiting customers' plot in Figure \ref{fig:waiting}. An important indicator for the customer satisfaction reported in Figure \ref{fig:detour-time-histogram} is the detour time, computed as the difference from the customer preferred travel time at trip destination and the actual arrival time. In the \Iotd\ service, fewer customers are affected by delays of more than $6$ minutes.

%
%

\subsection{MaaS demonstration}

In addition to the above simulation results, we further validated the feasibility of our architecture and approach in a real environment (Brainport) using real connected autonomous vehicles. The demonstration was carried out at the ITS Europe Congress, in Eindhoven, June 2019 \cite{ITS}. In the demonstration, two automated vehicles were used. A customer wanting to travel from the automotive campus of Helmond to Eindhoven booked a shared ride through our ride sharing service, which assigned one of the autonomous vehicles to the customer request. The vehicle, which was initially parked, drove itself autonomously from its parking location to the customer. This was operated by an automated valet parking (AVP) system developed by the AUTOPILOT project partners. The vehicle then picked up the customer and headed to Eindhoven. The ride sharing was also demonstrated in a broader context of mobility as a service, where platooning with the second automated vehicle was tested.

The demonstration was successfully repeated several times before and during the ITS Europe Congress. The preparation period increased our confidence in the feasibility of deploying the developed ridesharing service prototype onto the real World.
\section{Conclusion}

The paper proposed an IoT architecture for city-scale ridesharing services. The impact of traffic events affecting the vehicle travel times has been assessed via a simulation in a geographical area in the Netherlands. The results advocate for the need of IoT-aware ridesharing in urban environments, as an approach to improve the number of served customers as well as the utilization of the vehicles, and to limit the inconveniences experienced by customers, such as waiting time and detour time. The service has been successfully demonstrated during the ITS Europe Congress, 2019. Future work will focus on IoT simulations of larger scale.


\bibliography{ref}
\bibliographystyle{plain}

\end{document}